\documentclass[eqsecnum]{revtex4} 

\usepackage{latexsym} 
\usepackage{amsmath,amssymb} 

\newcommand{\XiT}{\xi_{(t)}} 
\newcommand{\XiI}{\xi_{(i)}} 
\newcommand{\XiH}{\xi_{(h)}} 
\newcommand{\BXiT}{\bar{\xi}_{(t)}} 
\newcommand{\BXiI}{\bar{\xi}_{(i)}} 
\newcommand{\BXiH}{\bar{\xi}_{(h)}} 
\newcommand{\OmI}{\Omega^{(i)}} 
\newcommand{\BOmI}{\bar{\Omega}^{(i)}} 
\newcommand{\JI}{J_{(i)}} 
\newcommand{\BJI}{\bar{J}_{(i)}} 
\newcommand{\Qcc}{{\cal Q}} 
\newcommand{\BQcc}{\bar{\cal Q}} 
\newcommand{\QspH}{Q_H} 
\newcommand{\BQspH}{\bar{Q}_H} 
\newcommand{\QspI}{Q_{\cal I}} 
\newcommand{\BQspI}{\bar{Q}_{\cal I}} 
\newtheorem{definition}{Definition} 

\begin{document} 
\title{ 
The first law of AdS black holes 
in higher curvature gravity} 

\author{Jun-ichirou Koga} 
\email{koga@gravity.phys.waseda.ac.jp} 
\affiliation{Advanced Research Institute for Science and Engineering, 
Waseda University, Shinjuku-ku, Tokyo 169-8555, Japan} 

\preprint{WU-AP/222/05}

\begin{abstract} 
We consider the first law of black hole thermodynamics 
in an asymptotically anti-de Sitter spacetime in the class of gravitational theories  
whose gravitational Lagrangian is an arbitrary function of the Ricci scalar. 
We first show that the conserved quantities in this class of gravitational theories 
constructed through conformal completion remain unchanged 
under the conformal transformation into the Einstein frame. 
We then prove that the mass and the angular momenta defined by 
these conserved quantities, along with the entropy defined by 
the Noether charge, satisfy the first law of black hole thermodynamics, 
not only in Einstein gravity but also in the higher curvature gravity 
within the class under consideration. 
We also point out that it is naturally understood in the symplectic formalism 
that the mass satisfying the first law should be necessarily defined 
associated with the timelike Killing vector nonrotating at infinity. 
Finally, a possible generalization into a wider class of gravitational theories 
is discussed. 
\end{abstract} 

\maketitle 


\section{Introduction} 

Although more than thirty years have passed since the laws of 
black hole thermodynamics were first formulated \cite{BardeenCH73}, 
it is quite recent that 
the thermodynamical variables of the Kerr--anti-de Sitter (AdS) black hole 
in Einstein gravity that satisfy the first law 
have been correctly identified \cite{GibbonsPP-}. 
Since asymptotically AdS spacetimes are attracting much attention \cite{AdSCFT} 
and black hole thermodynamics is expected to play a key role in quantum gravity, 
it is interesting to explore to what extent black hole thermodynamics is robust 
in an asymptotically AdS spacetime against quantum gravitational effects. 
A possible way to proceed in this direction is to consider 
the effects of higher curvature corrections to Einstein gravity, 
since such corrections naturally arise due to quantum effects 
\cite{CallanFMP85,BirrellDavies82}. 

In the case of an asymptotically flat black hole, 
black hole thermodynamics is endowed with a reliable foundation, in the sense that 
the first law has been shown to hold in a quite general context 
for a stationary and axisymmetric black hole with a bifurcate Killing horizon, 
without depending on particular solutions or gravitational theories \cite{FirstLawAF}.  
A bifurcate Killing horizon has a surface where the Killing vector that generates 
the Killing horizon vanishes, which is called a bifurcation surface, 
and a bifurcation surface is shown to be possessed 
by a generic regular Killing horizon with nonvanishing temperature \cite{RaczWald}. 
The first law of an asymptotically flat black hole 
is then proved \cite{FirstLawAF} in the symplectic formalism 
(the covariant phase space formalism) 
by employing Noether charges associated with diffeomorphism invariance. 
The Noether charge conjugate to the Killing vector that generates the Killing horizon 
gives the entropy of a black hole multiplied by its temperature 
when it is evaluated at the Killing horizon, 
and then one finds that the entropy is not necessarily given 
by one quarter of its area (in the units $c = G = \hbar = 1$) 
if higher curvature corrections are included. 
Thus, black hole entropy is defined based on the first law in the symplectic formalism. 
At spacial infinity, on the other hand, 
the Noether charge conjugate to the timelike Killing vector and those conjugate to 
the axial Killing vectors yield the Arnowitt--Deser--Misner mass 
and Komar's angular momenta, respectively. 
The latter fact is not affected by higher curvature corrections, 
since these corrections fall off rapidly enough at the asymptotically flat region   
and hence do not contribute to the mass and the angular momenta. 

In the case of an asymptotically AdS spacetime, however, 
black hole thermodynamics in the general context, 
as in the asymptotically flat case, has not been established yet. 
When we focus on Einstein gravity, the mass and the angular momenta of the Kerr--AdS 
black hole are shown to 
satisfy the first law of black hole thermodynamics \cite{GibbonsPP-}, 
if they are defined by the conserved quantities constructed 
by Ashtekar, Magnon, and Das 
through conformal completion \cite{AshtekarMagnonDas}, 
which we call conformal charges in this paper, while the mass should be 
defined associated with the timelike Killing vector nonrotating at infinity. 
In a more general context, but still within Einstein gravity, 
Hollands, Ishibashi and Marolf \cite{HollandsIM} have recently shown that 
Noether charges evaluated at infinity of an asymptotically AdS spacetime 
coincide with conformal charges. 
As we will see in this paper, 
this implies that the mass and the angular momenta of a black hole 
defined by these conformal charges, along with the entropy 
defined as one quarter of the area of the black hole horizon, 
satisfy the first law of black hole thermodynamics generally in Einstein gravity, 
not only for the Kerr--AdS black hole, under suitable assumptions 
on the asymptotic behavior of matter fields at infinity.   

Noether charges are actually defined 
by their variations in the phase space, not by Noether charges themselves.  
In order that Noether charges are integrable, one then needs to impose 
boundary conditions at infinity \cite{WaldZoupas00}, 
since otherwise Noether charges, which give the mass and the angular momenta, 
are not well-defined, which in turn invalidates the first law. 
While one might expect 
that the entropy of a black hole is always defined by the Noether charge 
on the horizon, not only for an asymptotically flat black hole but also for an asymptotically AdS black hole, 
defining entropy based on the first law does not make sense 
unless the first law itself is well-defined. 
It is thus very crucial that boundary conditions are imposed at infinity so that 
the mass and the angular momenta of a black hole are 
given by well-define Noether charges. 
On the other hand, boundary conditions 
at infinity are closely related with how an asymptotically AdS spacetime is defined. 
In order to define an asymptotically AdS spacetime in a rigorous manner, however, 
it is plausible to employ the conformal completion technique, with which 
we can construct well-defined conformal charges \cite{AshtekarMagnonDas}, 
i.e., integrated conserved quantities, not their variations, without subtracting 
background contribution or employing a counterterm.    
Then, it is remarkable that Noether charges at infinity have been proven to 
coincide with conformal charges in Einstein gravity \cite{HollandsIM}, 
since it implies that Noether charges that define  the mass and the angular momenta 
and satisfy the first law are indeed well-defined in Einstein gravity. 

In higher curvature gravity, an asymptotically AdS spacetime has been defined recently 
through the conformal completion technique and well-defined conformal charges 
have been explicitly constructed 
\cite{OkuyamaKoga05}.  
Then, it is interesting to investigate whether the mass and the angular momenta 
defined by these conformal charges satisfy the first law of 
an asymptotically AdS black hole, 
not only in Einstein gravity but also in higher curvature gravity. 
However, it is still unclear whether one can deal with arbitrary theories of gravity 
at a single stroke. 
We will thus mainly focus in this paper on the class 
of gravitational theories  
whose gravitational Lagrangian is an arbitrary function of the Ricci scalar alone, 
with the spacetime dimensionality greater than three. 
It is well-known that such theories of gravity 
can be conformally transformed into the Einstein frame. 
We will then begin in Sec. \ref{sec:TransAndConsQ} with a brief review 
on the definition of an asymptotically AdS spacetime and conformal charges  
in this class of gravitational theories. 
We will also analyze how these conformal charges 
behave under the conformal transformation into the Einstein frame. 
Based on the result in Sec. \ref{sec:TransAndConsQ}, we will investigate 
in Sec. \ref{sec:FirstLaw} whether the first law of black hole thermodynamics 
is satisfied when the mass and the angular momenta are defined by 
these conformal charges and the entropy is defined by the Noether charge on the horizon. 
In Sec. \ref{sec:ConcDis}, we will summarize this paper and 
discuss a possible generalization into a wider class of gravitational theories. 
Appendix is devoted to the simplest example, 
where we will calculate the mass and the angular momentum of the four-dimensional 
Kerr--AdS black hole by using the conformal completion technique 
in both the original frame and the Einstein frame. 

\section{Conformal transformation and conformal charges} 
\label{sec:TransAndConsQ} 

The action of the gravitational theories we consider here is 
given by 
\begin{equation} 
I = \int d^n x \sqrt{- g} \left[ \frac{f(R)}{16 \pi G} + L_m \right] , 
\label{eqn:ActionOrg} 
\end{equation} 
where $L_m$ is the Lagrangian of matter fields and the gravitational Lagrangian 
$f(R)$ is an arbitrary smooth function of the Ricci scalar $R$. 
The gravitational field equation that follows from the action Eq. (\ref{eqn:ActionOrg}) 
is given by 
\begin{equation}
f'(R) R_{a b} - \frac{1}{2} g_{a b} f(R) - \nabla_a \nabla_b f'(R) 
+ g_{a b} \nabla_c \nabla^c f'(R) = 8 \pi G T_{a b} ,
\label{eqn:GravEqO}
\end{equation} 
and thus a spacetime with the Ricci tensor $R_{a b}$ of the form 
\begin{equation} 
R_{a b} = \frac{R_0}{n} g_{a b} , 
\label{eqn:RicciConstCurv} 
\end{equation} 
is a solution to Eq. (\ref{eqn:GravEqO}),   
if the constant $R_0$ satisfies  
\begin{equation} 
{f'}^{- 1}(R_0) f(R_0) = \frac{2}{n} R_0 .  
\label{eqn:RicciEq} 
\end{equation} 
In particular, a spacetime with constant curvature possesses the Ricci tensor of the form of Eq. (\ref{eqn:RicciConstCurv}), 
and hence it describes the pure AdS spacetime 
with the curvature length $\ell$,   
if the constant $R_0$ is negative and so there exists a real constant $\ell$ 
such that 
\begin{equation} 
R_0 = - \frac{n ( n - 1 )}{\ell^2} .     
\label{eqn:RicciInf} 
\end{equation} 

A spacetime that asymptotically approaches the pure AdS spacetime 
is considered as an asymptotically AdS spacetime. 
By extending the definition in Einstein gravity \cite{AshtekarMagnonDas} 
and imposing the asymptotic condition on the Riemann tensor $R_{a b c d}$, 
which is regarded as ``a boundary condition at infinity'', 
an asymptotically AdS spacetime is defined in Ref. \cite{OkuyamaKoga05},  
so that it works also in higher curvature gravity, as 
\begin{definition} 
An $n$-dimensional spacetime $( {\cal M} , g_{a b} )$ with $n \geq 4$ 
is said to be an asymptotically AdS spacetime if there exists 
a spacetime $( \hat{\cal M} , \hat{g}_{a b} )$ 
with a boundary ${\cal I}$ equipped with the smooth metric $\hat{g}_{a b}$, 
which satisfies following conditions:
\begin{enumerate}
\item $\hat{\cal M} \setminus {\cal I} \cong {\cal M}$ 
and ${\cal I}\cong S^{n-2}\times \mathbb{R}$.
\item There exists a smooth scalar $\Omega$ on $\hat{\cal M}$, such that 
$\hat{g}_{a b} = \Omega^2 g_{a b}$ on ${\cal M}$, and 
$\Omega = 0$ and 
$\hat{\nabla}_a \Omega \neq 0$ on ${\cal I}$, 
where $\hat{\nabla}_a$ is the covariant derivative associated with 
the unphysical metric $\hat{g}_{a b}$. 
\item The physical metric $g_{a b}$ satisfies the gravitational equation on ${\cal M}$, and there exists a smooth tensor $\tau_{a b}$ such that the energy-momentum tensor 
$T_{a b}$ in a neighborhood of ${\cal I}$ is given by 
$T_{a b} = \Omega^{n-2} \tau_{a b}$.
\item There exist a smooth tensor $H_{abcd}$ and a real constant $\ell$ such that the Riemann tensor $R_{a b c d}$ associated with the physical metric $g_{a b}$ in a neighborhood of ${\cal I}$ is given by 
$R_{a b c d} + \ell^{- 2} \left( g_{a c} g_{d b} - g_{a d} g_{c b} \right) 
= \Omega^{n-5} H_{a b c d}$. 
\end{enumerate} 
\label{def:AsymAdS} 
\end{definition} 
In the case of four-dimensional spacetimes, however, 
we need to impose the so-called reflective boundary condition 
\cite{AshtekarMagnonDas} on the magnetic part of the Weyl tensor 
in order to obtain a universal group of asymptotic symmetries on ${\cal I}$. 
Although it will be plausible in future investigations 
to refine the definition of an asymptotically AdS spacetime 
to include in a natural manner this exceptional condition in four dimensions, 
here we simply follow the previous frameworks 
\cite{AshtekarMagnonDas,OkuyamaKoga05} 
and impose the reflective boundary condition in this paper. 
Then, it is shown that ${\cal I}$ is conformally flat and thus the asymptotic symmetry 
group on ${\cal I}$ is the $n$-dimensional AdS group. 
Furthermore, it has been shown in Ref. \cite{OkuyamaKoga05} that 
conformal charges $\Qcc[\xi]$ associated with this asymptotic symmetry group 
on ${\cal I}$ are given, in the class of gravitational theories 
Eq. (\ref{eqn:ActionOrg}), as  
\begin{equation}
\Qcc[\xi] 
= - \frac{\ell \, f'(R_0)}{8 \pi G (n-3)} \int_{\cal C} dx^{n-2} \sqrt{\hat{\sigma}} \:  
\hat{\cal E}_{a b} \: \xi^a \hat{N}^b , 
\label{eqn:QExprOrg}
\end{equation}
on an arbitrary cross section ${\cal C}$ of ${\cal I}$,   
where $\sqrt{\hat{\sigma}}$ is the volume element of ${\cal C}$, 
$\hat{N}^a$ is the future-pointing timelike unit normal to ${\cal C}$ 
(with respect to the unphysical metric $\hat{g}_{a b}$), 
and $\xi^a$ denotes the asymptotic Killing vectors, 
which generate the asymptotic symmetry group.  
The argument $R_0$ of $f'(R_0)$ is the value of the Ricci scalar $R$ on ${\cal I}$, 
which is given by Eq. (\ref{eqn:RicciInf}) in terms of the curvature length $\ell$ 
at infinity, and the tensor $\hat{{\cal E}}_{a b}$ is (the leading part of) 
the electric part of the Weyl tensor $C_{a b c d}$ associated with $g_{a b}$ defined by 
\begin{equation} 
\hat{{\cal E}}_{a b} \equiv \ell^2 \, \Omega^{5 - n} C_{a c b d} \: \hat{n}^c \hat{n}^d ,  
\label{eqn:EDefOr} 
\end{equation} 
where $\hat{n}^a$ is a normal to ${\cal I}$ defined as 
\begin{equation} 
\hat{n}_a \equiv \hat{\nabla}_a \Omega .  
\label{eqn:nDef} 
\end{equation} 
These conformal charges $\Qcc[\xi]$ are shown to 
satisfy a physically reasonable form of a balance equation. 
It is also important to bear in mind that the curvature length $\ell$ at infinity 
is determined 
once the form of the function $f(R)$ is given, and hence $\ell$ is fixed and 
cannot be varied continuously, 
even when a solution is deformed within the family of asymptotically AdS solutions. 

On the other hand, as it is well-known, the conformal transformation 
\begin{equation} 
g_{a b} \rightarrow \bar{g}_{a b} = \omega^2 g_{a b} ,  
\label{eqn:ConfTrOtoE} 
\end{equation} 
where the conformal factor $\omega$ is defined by 
\begin{equation} 
\omega \equiv \bigl( f'(R) \bigr)^{1 / ( n - 2 )} , 
\label{eqn:ConfFactToEin} 
\end{equation} 
and is assumed to be smooth and positive definite everywhere in $( {\cal M} , g_{a b} )$, 
except for spacetime singularities, 
transforms the physical spacetime $( {\cal M} , g_{a b} )$ in the original frame 
of the action Eq. (\ref{eqn:ActionOrg}) 
into the Einstein frame $( \bar{\cal M} , \bar{g}_{a b} )$.  
The action in the Einstein frame is given by 
\begin{equation} 
\bar{I} = \int d^n x \sqrt{- \bar{g}} \left[ \frac{1}{16 \pi G} \bigl\{ 
\bar{R} - \left( \bar{\nabla}_c \phi \right) \left( \bar{\nabla}^c \phi \right) 
- U(\phi) \bigr\} + \bigl( f'(R) \bigr)^{- n / ( n - 2 )} L_m \right] , 
\label{eqn:ActionEin} 
\end{equation} 
and the extra scalar field $\phi$ is defined as 
\begin{equation} 
\phi = \ln \bigl( f'(R) \bigr)^{\sqrt{( n - 2 ) / ( n - 1 )}} . 
\label{eqn:ExtraScalarDef} 
\end{equation} 
In the Einstein frame, 
the Ricci scalar $R$ associated with the original metric $g_{a b}$ 
is considered as a function of $\phi$, which is implicitly  defined by 
Eq. (\ref{eqn:ExtraScalarDef}), and is denoted by ${\cal R}$. 
The potential $U(\phi)$ of $\phi$ is then found to be given  
in terms of ${\cal R}$ as
\begin{equation} 
U(\phi) = \bigl( f'({\cal R}) \bigr)^{- 2 / ( n - 2 )} \left[ 
{\cal R} - \bigl( f'({\cal R}) \bigr)^{- 1} f({\cal R}) \right] . 
\label{eqn:UDef} 
\end{equation} 
The field equations in the Einstein frame, 
apart from those for the ordinary matter fields, are derived as 
\begin{eqnarray} 
& & \bar{R}_{a b} - \frac{1}{2} \bar{g}_{a b} \bar{R} + \bar{\Lambda} \bar{g}_{a b} 
= 8 \pi G \bar{T}_{a b} , 
\label{eqn:GEqE} \\ 
& & 2 \bar{\nabla}^c \bar{\nabla}_c \phi - U'(\phi) 
= 16 \pi G \frac{n \bigl( f'({\cal R}) \bigr)^{- n / ( n - 2 )}}{\sqrt{( n - 1 ) ( n - 2 )}}  
L_m . 
\label{eqn:ScEqE} 
\end{eqnarray} 
In Eq. (\ref{eqn:GEqE}), we subtracted out a possible cosmological term 
$- \bar{\Lambda} \bar{g}_{a b}$ 
from the matter fields, and thus the energy-momentum tensor $\bar{T}_{a b}$ 
in the Einstein frame is defined by 
\begin{equation} 
\bar{T}_{a b} \equiv \bigl( f'({\cal R}) \bigr)^{- 1} T_{a b} + \bar{T}^{(\phi)}_{a b} 
+ \frac{\bar{\Lambda}}{8 \pi G} \bar{g}_{a b} , 
\label{eqn:TbarDef} 
\end{equation} 
where $\bar{\Lambda}$ is an effective cosmological constant,  
which is to be determined below,  
and the energy-momentum tensor $\bar{T}^{(\phi)}_{a b}$ of the extra scalar filed 
$\phi$ is given by  
\begin{equation} 
\bar{T}^{(\phi)}_{a b} = \frac{1}{8 \pi G} \left\{ \left( \bar{\nabla}_a \phi \right) 
\left( \bar{\nabla}_b \phi \right) - \frac{1}{2} \bar{g}_{a b} 
\left[ \left( \bar{\nabla}^c \phi \right) \left( \bar{\nabla}_c \phi \right) 
+ U(\phi) \right] \right\} .  
\label{eqn:EMTExSc} 
\end{equation} 

If the physical spacetime $( {\cal M} , g_{a b} )$ in the original frame is the pure 
AdS spacetime (with $T_{a b} = 0$), 
the Ricci scalar $R$ associated with the physical metric $g_{a b}$ in the original frame 
is given by the constant $R_0$ that satisfies Eq. (\ref{eqn:RicciEq}), 
and thus it is denoted by ${\cal R}_0$ in the Einstein frame. 
The conformal factor $\omega$ and the extra scalar 
field $\phi$ are then given by the constants $\omega_0$ and $\phi_0$, 
which are defined by Eqs. (\ref{eqn:ConfFactToEin}) 
and (\ref{eqn:ExtraScalarDef}), respectively, with $R = R_0$ substituted. 
In this case, the energy-momentum tensor $\bar{T}^{(\phi)}_{a b}$ of 
the extra scalar field $\phi$ is given by 
\begin{equation} 
\bar{T}^{(\phi)}_{a b} = - \frac{1}{16 \pi G} \, \bar{g}_{a b} \, U(\phi_0) 
= - \frac{1}{8 \pi G} \frac{n - 2}{2 n} \bar{g}_{a b} 
\bigl( f'({\cal R}_0) \bigr)^{- 2 / ( n - 2 )} {\cal R}_0 .    
\end{equation} 
By defining the effective cosmological constant $\bar{\Lambda}$ as 
\begin{equation} 
\bar{\Lambda} \equiv 
\frac{n - 2}{2 n} \bigl( f'({\cal R}_0) \bigr)^{- 2 / ( n - 2 )} {\cal R}_0 ,   
\end{equation} 
we then see that $\bar{T}^{(\phi)}_{a b}$ is written as 
\begin{equation} 
\bar{T}^{(\phi)}_{a b} = - \frac{1}{8 \pi G} \, \bar{\Lambda} \, \bar{g}_{a b} , 
\label{eqn:TBarPhiAdS} 
\end{equation} 
and hence $\bar{T}_{a b}$ vanishes. 
We also notice that the conformal transformation Eq. (\ref{eqn:ConfTrOtoE}) 
in this case reduces to rescaling of the metric. If we employ the standard static 
chart of the pure AdS spacetime, where the metric $g_{a b}$ in the original frame 
is given by 
\begin{equation} 
d s^2 = - \left( 1 + \frac{r^2}{\ell^2} \right) d t^2 
+ \left( 1 + \frac{r^2}{\ell^2} \right)^{- 1} d r^2 + r^2 d \Omega^{n - 2} , 
\label{eqn:StaticPureAdSO} 
\end{equation} 
the metric $\bar{g}_{a b}$ in the Einstein frame is written as 
\begin{equation} 
d \bar{s}^2 = - \left( 1 + \frac{\bar{r}^2}{\bar{\ell}^2} \right) d \bar{t}^2 
+ \left( 1 + \frac{\bar{r}^2}{\bar{\ell}^2} \right)^{- 1} d \bar{r}^2 
+ \bar{r}^2 d \Omega^{n - 2} , 
\label{eqn:StaticPureAdSE} 
\end{equation} 
i.e., the same form as Eq. (\ref{eqn:StaticPureAdSO}), 
while the quantities of the dimension of length are rescaled as 
\begin{equation} 
\bar{t} \equiv \omega_0 t , ~~~~~ \bar{r} \equiv \omega_0 r , ~~~~~ 
\bar{\ell} \equiv \omega_0 \ell .  
\label{eqn:Rescaling} 
\end{equation} 

We now consider that the physical spacetime $( {\cal M} , g_{a b} )$ 
in the original frame Eq. (\ref{eqn:ActionOrg}) is an asymptotically AdS spacetime, 
and hence there exists a conformally completed spacetime 
$( \hat{\cal M} , \hat{g}_{a b} )$, which satisfies 
all of the four conditions of Definition \ref{def:AsymAdS}. 
We then note that 
there exists a conformal transformation that brings 
the physical spacetime $( \bar{\cal M} , \bar{g}_{a b} )$ in the Einstein frame 
into the spacetime $( \hat{\cal M} , \hat{g}_{a b} )$, i.e.,  
the identical spacetime to that obtained from conformal completion of 
the physical spacetime $( {\cal M} , g_{a b} )$ in the original frame by 
$g_{a b} \rightarrow \hat{g}_{a b} = \Omega^2 g_{a b}$. 
Actually, we see, by setting as $\bar{\Omega} = \omega^{- 1} \Omega$, 
that the conformal transformation 
\begin{equation} 
\bar{g}_{a b} \rightarrow \hat{g}_{a b} = \bar{\Omega}^2 \bar{g}_{a b} 
\label{eqn:ConfCompEinPr} 
\end{equation} 
transforms the physical spacetime $( \bar{\cal M} , \bar{g}_{a b} )$ 
in the Einstein frame into $( \hat{\cal M} , \hat{g}_{a b} )$, and that 
$\bar{\Omega} = 0$ and $\hat{\nabla}_a \bar{\Omega} \neq 0$ on ${\cal I}$, 
because $\omega$ is assumed to be smooth and nonvanishing. 
In addition, since the fourth condition of Definition \ref{def:AsymAdS} 
is satisfied by the physical spacetime $( {\cal M} , g_{a b} )$ in the original frame, 
${\cal R}$ possesses the asymptotic form 
\begin{equation} 
{\cal R} = {\cal R}_0 + \bar{\Omega}^{n - 1} {\cal R}_1 , 
\end{equation} 
and correspondingly the conformal factor $\omega$ 
and the extra scalar field $\phi$ are written near ${\cal I}$ as 
\begin{equation} 
\omega = \omega_0 + \bar{\Omega}^{n - 1} \omega_1 , ~~~~~~ 
\phi = \phi_0 + \bar{\Omega}^{n - 1} \phi_1 , 
\label{eqn:OmegaPhiAsym} 
\end{equation} 
where ${\cal R}_1$, $\omega_1$ and $\phi_1$ 
are smooth functions on ${\cal I}$. 
Since we also have $U'(\phi_0) = 0$, we find from 
Eqs. (\ref{eqn:EMTExSc}) and (\ref{eqn:OmegaPhiAsym}) 
that $\bar{T}^{(\phi)}_{a b}$ is given as 
\begin{equation} 
\bar{T}^{(\phi)}_{a b} = - \frac{1}{8 \pi G} \, \bar{\Lambda} \, \bar{g}_{a b} 
+ O(\bar{\Omega}^{2 ( n - 2 )}) , 
\label{eqn:EneMomEAsym} 
\end{equation} 
which implies that $\bar{T}^{(\phi)}_{a b}$ approaches Eq. (\ref{eqn:TBarPhiAdS}) 
faster than the rate of $\bar{\Omega}^{n - 2}$.  
Furthermore, the energy-momentum tensor $T_{a b}$ in the original frame 
falls off as $\Omega^{n - 2}$ due to the third condition of 
Definition \ref{def:AsymAdS} and $(f'({\cal R}))^{- 1}$ is smooth at ${\cal I}$. 
We therefore see that $(f'({\cal R}))^{- 1} T_{a b}$ and hence 
the energy momentum tensor $\bar{T}_{a b}$ 
in the Einstein frame fall off at the rate of $\bar{\Omega}^{n - 2}$. 
Finally, from Eq. (\ref{eqn:OmegaPhiAsym}) and the fourth condition of 
Definition \ref{def:AsymAdS}, 
along with the relation between 
the Riemann tensor $R_{a b c d}$ associated with $g_{a b}$ 
and $\bar{R}_{a b c d}$ associated with $\bar{g}_{a b}$,  
we can show that there exist a smooth tensor 
$\bar{H}_{a b c d}$ and a real constant $\bar{\ell}$ such that 
\begin{equation} 
\bar{R}_{a b c d} + \frac{1}{\bar{\ell}^2} 
\left( \bar{g}_{a c} \bar{g}_{d b} - \bar{g}_{a d} \bar{g}_{c b} \right) 
= \bar{\Omega}^{n - 5} \bar{H}_{a b c d} , 
\label{eqn:AsymAdSCond4} 
\end{equation} 
where the curvature length $\bar{\ell}$ at infinity in the Einstein frame 
is shown to be related with $\ell$ in the original frame 
as described by Eq. (\ref{eqn:Rescaling}). 
Therefore, we see that all of the four conditions of Definition \ref{def:AsymAdS}  
are satisfied by the physical spacetime $( \bar{\cal M} , \bar{g}_{a b} )$ 
in the Einstein frame, and hence 
$( \bar{\cal M} , \bar{g}_{a b} )$ is an asymptotically AdS spacetime, 
if so is the physical spacetime $( {\cal M} , g_{a b} )$ in the original frame. 
The asymptotic infinity of $( \bar{\cal M} , \bar{g}_{a b} )$ then 
corresponds to the boundary $\bar{\cal I}$ of $( \hat{\cal M} , \hat{g}_{a b} )$ 
where $\bar{\Omega}$ vanishes. 
However, we note here that ${\cal I}$ in the original frame 
and $\bar{\cal I}$ in the Einstein frame are actually the same boundary  
of the same spacetime $( \hat{\cal M} , \hat{g}_{a b} )$. 
While one can choose different conformal 
completion \cite{AshtekarMagnonDas} by choosing 
$\bar{\Omega}$ different from $\omega^{- 1} \Omega$, 
there thus exists a mapping with which we can identify 
an arbitrary cross section $\bar{\cal C}$ of $\bar{\cal I}$ 
in the Einstein frame
with the corresponding cross section ${\cal C}$ of ${\cal I}$ in the original frame. 

Once the physical spacetime $( \bar{\cal M} , \bar{g}_{a b} )$ in the Einstein frame 
is found to be asymptotically AdS, we need not persist in setting as 
$\bar{\Omega} = \omega^{- 1} \Omega$, as we mentioned just above.  
We then let $\bar{\Omega}$ be given by 
\begin{equation} 
\bar{\Omega} \equiv \varpi \, \Omega ,  
\label{eqn:omegaPrimeDef} 
\end{equation} 
where $\varpi$ is an arbitrary smooth function that does not vanish on ${\cal I}$, 
and consider the conformal transformation 
\begin{equation} 
\bar{g}_{a b} \rightarrow \tilde{g}_{a b} = \bar{\Omega}^2 \bar{g}_{a b} ,   
\label{eqn:ConfCompE} 
\end{equation} 
which transforms the physical spacetime $( \bar{\cal M} , \bar{g}_{a b} )$ 
in the Einstein frame into a conformally completed spacetime 
$( \tilde{\cal M} , \tilde{g}_{a b} )$. 
Then, conformal charges $\BQcc[\xi]$ on a cross section $\bar{\cal C}$ 
of $\bar{\cal I}$ in the Einstein frame are given \cite{AshtekarMagnonDas} by 
\begin{equation}
\BQcc[\xi] 
= - \frac{\bar{\ell}}{8\pi G (n-3)} \int_{\bar{\cal C}} dx^{n-2} \sqrt{\tilde{\sigma}} \: 
\tilde{{\cal E}}_{a b} \: \xi^a \tilde{N}^b , 
\label{eqn:QExprEin}
\end{equation}
where $\sqrt{\tilde{\sigma}}$ is the volume element of $\bar{\cal C}$, 
$\tilde{N}^a$ is the future-pointing timelike unit normal to $\bar{\cal C}$ (with respect 
to $\tilde{g}_{a b}$), $\tilde{{\cal E}}_{a b}$ is defined as   
\begin{equation} 
\tilde{{\cal E}}_{a b} 
\equiv \bar{\ell}^2 \: \bar{\Omega}^{5 - n} \bar{C}_{a c b d} \: \tilde{n}^c \tilde{n}^d , 
\label{eqn:EDefEin} 
\end{equation} 
and $\tilde{n}_c$ is given by 
\begin{equation} 
\tilde{n}_c \equiv \tilde{\nabla}_c \bar{\Omega} . 
\label{eqn:TildenDef} 
\end{equation} 
The conformal charges $\BQcc[\xi]$ in the Einstein frame 
also satisfy the balance equation 
of the same form as that in the original frame. 
We note also that the asymptotic Killing vectors $\xi^a$ on ${\cal I}$ 
in the original frame are asymptotic Killing vectors on $\bar{\cal I}$  
in the Einstein frame, as well.  

Now we compare the conformal charges $\BQcc[\xi]$ 
in the Einstein frame with $\Qcc[\xi]$ in the original frame. 
We first notice from Eqs. (\ref{eqn:nDef}) and (\ref{eqn:TildenDef}) 
that $\tilde{n}^a$ is related with $\hat{n}^a$ as 
$\tilde{n}^a = \varpi^{- 1} \omega_0^{- 2} \hat{n}^a$ on ${\cal I}$, 
and then we obtain 
$\tilde{\cal E}_{a b} = \varpi^{3 - n} \hat{\cal E}_{a b}$ on ${\cal I}$ 
by using Eq. (\ref{eqn:Rescaling}). 
We can also show $\tilde{N}^a = \varpi^{- 1} \omega_0^{- 1} \hat{N}^a$ and 
$\sqrt{\tilde{\sigma}} = \varpi^{n - 2} \omega_0^{n - 2} \sqrt{- \hat{\sigma}}$. 
By substituting these relations into Eq. (\ref{eqn:QExprEin}) 
and utilizing the mapping from $\bar{\cal C}$ to ${\cal C}$, 
we find that the conformal charges   
$\BQcc[\xi]$ in the Einstein frame are written as 
\begin{equation} 
\BQcc[\xi] 
= - \frac{\ell \; \omega_0^{n - 2}}{8\pi G (n-3)} 
\int_C dx^{n-2} \sqrt{\hat{\sigma}} \: \hat{{\cal E}}_{a b} \: \xi^a \hat{N}^b .  
\label{eqn:barQTransPre} 
\end{equation} 
It is interesting to notice here that 
the conformal charges depend on the conformal factor $\omega$, whereas they 
do not on $\varpi$, as they should not \cite{AshtekarMagnonDas}. 

From Eqs. (\ref{eqn:QExprOrg}), (\ref{eqn:ConfFactToEin}), and (\ref{eqn:barQTransPre}), 
we therefore find 
\begin{equation} 
\BQcc[\xi] = \Qcc[\xi] ,  
\label{eqn:QEqual} 
\end{equation} 
i.e., the conformal charges $\Qcc[\xi]$ in the original frame and 
$\BQcc[\xi]$ in the Einstein frame coincide with each other 
{\it for the same asymptotic Killing vector} $\xi^a$. 

\section{The first law} 
\label{sec:FirstLaw} 

We consider in this section the first law of black hole thermodynamics 
in an asymptotically AdS black hole spacetime in both the original frame 
and the Einstein frame. 
We assume here, however, that global charges 
independent of the mass and the angular momenta, 
such as the electromagnetic charge, are not possessed by the black hole, 
and that the matter fields fall off rapidly enough so that they not only satisfy 
the third condition of Definition \ref{def:AsymAdS}, but also make no contribution 
to Noether charges at infinity, which is actually shown to be true 
in Ref. \cite{HollandsIM} under suitable assumptions on the matter fields. 

We then consider  
a stationary and axisymmetric black hole spacetime with a bifurcate Killing horizon 
in the original frame.  
Thus, there exist a timelike Killing vector $\XiT^a$ and axial Killing vectors  
$\XiI^a$ in the original frame, and the linear combination $\XiH^a$ 
of $\XiT^a$ and $\XiI^a$ defined by 
\begin{equation} 
\XiH^a \equiv \XiT^a + \OmI \XiI^a , 
\label{eqn:HorGenODef} 
\end{equation} 
which generates the Killing horizon, vanishes 
on the bifurcation surface, where $\OmI$ are the angular velocities of 
the black hole. (There exist $[ ( n - 1) / 2  ]$ axial Killing vectors. 
The index $(i)$ labels these axial Killing vectors and 
the corresponding angular velocities, 
and hence it runs from $1$ to $[ ( n - 1) / 2  ]$. 
Summation over the index $(i)$ is also understood in Eq. (\ref{eqn:HorGenODef}).) 
However, the Killing vectors $\XiT^a$ and $\XiI^a$ are not uniquely determined, 
because any linear combinations of them with constant coefficients 
are also Killing vectors. 
In addition, as it has been pointed out by Gibbons et al. \cite{GibbonsPP-}, 
one has to take care that the time coordinate nonrotating at infinity 
with respect to ``the AdS background'' is used to define 
the mass of an asymptotically AdS black hole. 
In a coordinate system where the metric $g_{a b}$ behaves asymptotically as 
Eq. (\ref{eqn:StaticPureAdSO}), for example, 
with which one can naturally define ``the AdS background'', 
the mass should be defined associated with the time coordinate $t$. 
We therefore set $\XiT^a$ and $\XiI^a$ so that all of them are 
orthogonal to each other on ${\cal I}$ 
and the orbits of $\XiI^a$ on ${\cal I}$ close with the period of $2 \pi$. 
(See also the comment below, where we discuss that the Killing vectors should be 
necessarily fixed with respect to ``the AdS background'' in the symplectic formalism.) 

We note here that the conformal factor $\omega$ is 
a function only of the Ricci scalar $R$ 
and the Lie derivatives of $R$ along the Killing vectors $\XiT^a$ and $\XiI^a$ vanish, 
which implies that the Lie derivatives of the physical metric $\bar{g}_{a b}$ 
in the Einstein frame along these vectors vanish, as well. 
Therefore, there exist Killing vectors $\BXiT^a$ and 
$\BXiI^a$ also in the Einstein frame, which are given by constant multiples of 
$\XiT^a$ and $\XiI^a$, respectively. 
We then normalize 
the axial Killing vectors $\BXiI^a$ as 
\begin{equation} 
\XiI^a = \BXiI^a .    
\label{eqn:XiIRel} 
\end{equation} 
On the other hand, 
we set the relative normalization between 
$\XiT^a$ and $\BXiT^a$ as 
\begin{equation} 
\BXiT^a = \omega_0^{- 1} \XiT^a , 
\label{eqn:XiTRel} 
\end{equation} 
in accordance with the rescaling property described by Eq. (\ref{eqn:Rescaling}). 
In a coordinate system that behaves asymptotically as Eq. (\ref{eqn:StaticPureAdSO}), 
we then normalize the timelike Killing vector $\XiT^a$ in the original frame as 
$\XiT^a = \left( \partial / \partial t \right)^a$,  
and thus the timelike Killing vector $\BXiT^a$ in the Einstein frame is 
expressed as 
$\BXiT^a = \left( \partial / \partial \bar{t} \right)^a$, 
where $\bar{t}$ is the time coordinate in Eq. (\ref{eqn:StaticPureAdSE}). 

Since the Killing vectors $\XiT^a$ and $\XiI^a$, 
and hence $\BXiT^a$ and $\BXiI^a$, are asymptotic Killing vectors, 
we can define the conformal charges $\Qcc[\XiT]$ and $\Qcc[\XiI]$ 
associated with $\XiT^a$ and $\XiI^a$ in the original frame $( {\cal M} , g_{a b} )$, 
and $\BQcc[\BXiT]$ and $\BQcc[\BXiI]$ associated with 
$\BXiT^a$ and $\BXiI^a$ in the Einstein frame $( \bar{\cal M} , \bar{g}_{a b} )$.  
By using Eqs. (\ref{eqn:QEqual}), (\ref{eqn:XiIRel}), and (\ref{eqn:XiTRel}),  
as well as the fact that $\Qcc[\xi]$ is linear in $\xi^a$ 
for arbitrary asymptotic Killing vectors $\xi^a$, we then obtain  
\begin{equation} 
\Qcc[\XiT] = \omega_0 \BQcc[\BXiT] , ~~~~~ 
\Qcc[\XiI] = \BQcc[\BXiI] . 
\label{eqn:QXiRels} 
\end{equation} 
The mass $\bar{M}$ and the angular momenta $\BJI$ in the Einstein frame 
are defined, as if in Einstein gravity,  
by the conformal charges $\Qcc[\xi]$ as 
\begin{equation} 
\bar{M} \equiv \BQcc[\BXiT] , ~~~~~ 
\BJI \equiv - \BQcc[\BXiI] .  
\label{eqn:MassAngMomDefE} 
\end{equation} 
We will show in Appendix that Eq. (\ref{eqn:MassAngMomDefE}) correctly yields 
the mass and the angular momentum of the four-dimensional Kerr--AdS black hole 
in Einstein gravity \cite{GibbonsPP-}, 
while the sign of the second equation in Eq. (\ref{eqn:MassAngMomDefE}) is 
opposite to that presented in Ref. \cite{GibbonsPP-}. 
(See the comment at the end of Appendix below.) 
We then define the mass $M$ and the angular momenta $\JI$ of the black hole 
in the original frame also by 
\begin{equation} 
M \equiv \Qcc[\XiT] , ~~~~~ 
\JI \equiv - \Qcc[\XiI] . 
\label{eqn:MassAngMomDefO} 
\end{equation} 
Actually, one can show that the definition of the mass 
in Eq. (\ref{eqn:MassAngMomDefO}) reproduces 
the thermodynamical energy derived in Ref. \cite{NojiriOdintsov02}  
for spherically symmetric black holes in the class of gravitational theories  
Eq. (\ref{eqn:ActionOrg}). 
(It has been shown that Eq. (\ref{eqn:MassAngMomDefO}) 
gives the correct mass also in Einstein--Gauss--Bonnet gravity \cite{OkuyamaKoga05}.) 
From Eqs. (\ref{eqn:QXiRels}), 
(\ref{eqn:MassAngMomDefE}) and (\ref{eqn:MassAngMomDefO}), 
we therefore find that the masses and the angular momenta 
in the two frames are related as   
\begin{equation} 
M = \omega_0 \bar{M} , ~~~~~ \JI = \BJI . 
\label{eqn:MassAngMomRel} 
\end{equation} 

Now we recall the fact that a conformal transformation does not 
alter the causal structure of a spacetime. 
In addition, we can show that there exists a Killing vector,  
which becomes null on the hypersurface in the Einstein frame 
that corresponds to the Killing horizon in the original frame. 
Indeed, by defining the angular velocities $\BOmI$ in the Einstein frame by 
\begin{equation} 
\BOmI \equiv \omega_0^{- 1} \OmI , 
\label{eqn:AngVelRel} 
\end{equation} 
we find that the Killing vector $\BXiH^a$ defined by 
\begin{equation} 
\BXiH^a \equiv \BXiT^a + \BOmI \BXiI^a 
= \omega_0^{- 1} \XiH^a , 
\label{eqn:HorGenEDef} 
\end{equation} 
is null on the hypersurface in the Einstein frame where 
the Killing vector $\XiH^a$ defined by Eq. (\ref{eqn:HorGenODef}) 
becomes null in the original frame. 
Moreover, as we see from Eqs. (\ref{eqn:HorGenODef}) and (\ref{eqn:HorGenEDef}), 
the bifurcation surface in the original frame is a bifurcation surface 
also in the Einstein frame, because $\BXiH^a$ vanishes when $\XiH^a$ does. 
Therefore, the physical spacetime $( \bar{\cal M} , \bar{g}_{a b} )$ 
in the Einstein frame is found to be 
a black hole spacetime with a bifurcate Killing horizon, 
if so is the physical spacetime $( {\cal M} , g_{a b} )$ in the original frame, 
and the bifurcation surface appears at the same location in both the frames. 
We also find that the rescaling of the timelike Killing vector Eq. (\ref{eqn:XiTRel}) 
results in the rescaling of the surface gravity of the black hole. 
The surface gravity $\kappa$ in the original frame and $\bar{\kappa}$ 
in the Einstein frame, which are defined by 
\begin{equation} 
\XiT^b \nabla_b \XiT^a \equiv \kappa \: \XiT^a , ~~~~~~ 
\BXiT^b \bar{\nabla}_b \BXiT^a \equiv \bar{\kappa} \: \BXiT^a  
\label{eqn:SurfGravDef} 
\end{equation} 
on the horizon, are related as  
\begin{equation} 
\bar{\kappa} = \omega_0^{- 1} \kappa , 
\label{eqn:SurfGravRel} 
\end{equation} 
where $\nabla_b$ and $\bar{\nabla}_b$ 
are the covariant derivatives associated with $g_{a b}$ and $\bar{g}_{a b}$, 
respectively. 

One may expect that black hole entropy is defined without being affected 
by the geometric structure at infinity, since it is evaluated quasi-locally 
only on the horizon. This suggests that the entropy of 
an asymptotically AdS black hole is defined in the same manner as 
that of an asymptotically flat black hole \cite{FirstLawAF}. 
Thus, by using the Noether charge conjugate to $\XiH^a$ 
evaluated at the bifurcation surface $H$, which we denote by $\QspH[\XiH]$, 
we define the entropy $S$ of the black hole in the original frame by  
\begin{equation} 
\QspH[\XiH] \equiv \frac{\kappa}{2 \pi} S , 
\label{eqn:EntropyODef} 
\end{equation} 
where $\QspH[\XiH]$ is calculated from the action 
Eq. (\ref{eqn:ActionOrg}) as 
\begin{equation} 
\QspH[\XiH] = \frac{\kappa}{8 \pi G} \int_H d x^{n - 2}  \sqrt{\sigma} \: f'(R) . 
\label{eqn:NoetherHO} 
\end{equation} 
In the case of an asymptotically flat black hole, it has been proven 
\cite{JacobsonKM95,KogaMaeda98} that 
the entropy of a black hole in the original frame and that in the Einstein frame 
coincide with each other. 
It occurs essentially because the Noether charge $\QspH[\XiH]$ in the original frame 
coincides with $\BQspH[\XiH]$ in the Einstein frame for 
the same Killing vector $\XiH^a$. 
Since the Noether charge $\QspH[\XiH]$ depends only on tensors at the horizon,   
we can repeat exactly the same argument also in the case of 
an asymptotically AdS black hole.  
Actually, from Eqs. (\ref{eqn:ConfTrOtoE}) and (\ref{eqn:ConfFactToEin}), we obtain 
\begin{equation} 
\BQspH[\XiH] = \frac{\kappa}{8 \pi G} \int_H d x^{n - 2}  \sqrt{\bar{\sigma}} 
= \frac{\kappa}{8 \pi G} \int_H d x^{n - 2}  \sqrt{\sigma} \: f'(R) ,  
\label{eqn:NoetherHE} 
\end{equation} 
and hence we have 
\begin{equation} 
\QspH[\XiH] = \BQspH[\XiH] 
\label{eqn:NoetherEqual} 
\end{equation} 
also for an asymptotically AdS black hole. 
However, as we see from Eq. (\ref{eqn:SurfGravDef}), 
the surface gravity $\bar{\kappa}$ in the Einstein frame 
is defined with respect to $\BXiH^a$, not $\XiH^a$, and therefore 
the entropy $\bar{S}$ in the Einstein frame 
should be defined by the Noether charge $\BQspH[\BXiH]$ conjugate to $\BXiH^a$ as  
\begin{equation} 
\BQspH[\BXiH] \equiv \frac{\bar{\kappa}}{2 \pi} \bar{S} . 
\label{eqn:EntropyEDef} 
\end{equation} 
Since the Noether charge $\BQspH[\BXiH]$ is linear in $\BXiH^a$, 
we see from Eqs. (\ref{eqn:XiTRel}), (\ref{eqn:SurfGravRel}), 
(\ref{eqn:EntropyODef}), (\ref{eqn:NoetherEqual}), and (\ref{eqn:EntropyEDef}) 
that the entropy $S$ in the original frame and $\bar{S}$ 
in the Einstein frame coincide with each other, i.e.,  
\begin{equation} 
\bar{S} = S . 
\label{eqn:EntropyEqual} 
\end{equation} 

For an arbitrary gravitational theory with diffeomorphism invariance, 
it is shown in the symplectic formalism \cite{FirstLawAF} that 
the variations of Noether charges conjugate to arbitrary Killing vectors $\xi^a$ 
satisfy the spacial conservation law as 
\begin{equation} 
\delta \QspH[\xi] = \delta \QspI[\xi] , 
\label{eqn:NoetherConserv} 
\end{equation} 
where $\QspI[\xi]$ denotes the Noether charges evaluated at infinity. 
It is crucial here to notice that Eq. (\ref{eqn:NoetherConserv}) is valid whatever 
the asymptotic structure of the spacetime at infinity is, 
as long as the Noether charges are well-defined. 
Then, if the Noether charges are indeed integrable, 
the first law of black hole thermodynamics 
\begin{equation} 
\frac{\kappa}{2 \pi} \delta S = \delta M - \OmI \delta \JI 
\end{equation} 
follows by setting as $\xi^a = \XiH^a$ and defining the mass $M$, 
the angular momenta $\JI$, and the entropy $S$ by 
\begin{equation} 
M \equiv \QspI[\XiT] , ~~~~~ 
\JI \equiv - \QspI[\XiI] , ~~~~~ 
S \equiv \frac{2 \pi}{\kappa} \QspH[\XiH] , 
\label{eqn:ThemVarDefSymp} 
\end{equation} 
where we note that the variations of 
the surface gravity vanish when evaluated on the bifurcation surface 
and that $\delta \QspI[\XiH]$ is linear in $\XiH^a$. 

It is worth mentioning here that Killing vectors $\xi^a$, to which Noether charges 
are conjugate, must be kept fixed under the variations 
that appear in Eq. (\ref{eqn:NoetherConserv}). 
This is because these variations are actually identical to those 
in the variational principle of the action. 
Therefore, they act only on dynamical fields, such as the metric and matter fields, 
but should not act on the Killing vectors. 
On the other hand, these variations are subject to 
boundary conditions at infinity, by definition.  
(Recall the fact that boundary conditions should be imposed so that 
Noether charges are ensured to be integrable. This implies that 
the variations of Noether charges are subject to the boundary conditions, 
since otherwise Noether charges are not well-defined.) 
These facts imply that the Killing vectors $\xi^a$ should remain fixed with respect to 
the universal structure determined by the boundary conditions at infinity. 
In the case of an asymptotically AdS black hole, in particular, 
one should necessarily use the timelike Killing vector that is 
fixed with respect to ``the AdS background'' 
in order to define the mass by Eq. (\ref{eqn:ThemVarDefSymp}). 
Thus, it is naturally understood in the symplectic formalism 
why the mass satisfying the first law must be defined associated with 
the timelike Killing vector nonrotating at infinity. 

In Einstein gravity, it has been also shown that 
Noether charges evaluated at infinity of an asymptotically AdS spacetime 
coincide with conformal charges \cite{HollandsIM}. 
More precisely, we have
\begin{equation} 
\delta \Qcc[\xi] = \delta \QspI[\xi] , 
\label{eqn:ConfNoetEqual} 
\end{equation} 
for arbitrary asymptotic Killing vectors $\xi^a$, 
when matter fields fall off rapidly enough at infinity. In particular, 
a scalar field with the canonical kinetic term is shown \cite{HollandsIM} 
to make no contribution to $\delta \QspI[\xi]$ if it falls off as $\Omega^{( n - 1 ) / 2}$. 
We emphasize here that Eq. (\ref{eqn:ConfNoetEqual}) implies that Noether charges 
$\QspI[\xi]$ are well-defined, since conformal charges $\Qcc[\xi]$ are integrable 
by definition. 
One may notice, however, that the definition of an asymptotically AdS spacetime 
in Ref. \cite{HollandsIM} differs from Definition \ref{def:AsymAdS} in this paper. 
Actually, an asymptotically AdS spacetime is defined in Ref. \cite{HollandsIM} 
by imposing the condition that the induced metric on ${\cal I}$ is conformal 
to that of the Einstein static universe, instead of imposing an asymptotic condition 
on the curvature tensor. However, as we mentioned before, the asymptotic conditions 
of Definition \ref{def:AsymAdS}, along with the reflective boundary condition 
in four dimensions, imply that ${\cal I}$ is conformally flat. Since a conformally 
flat spacetime is conformal to the Einstein static universe, we see that 
Eq. (\ref{eqn:ConfNoetEqual}) indeed holds under the conditions of 
Definition \ref{def:AsymAdS}. 

In the Einstein frame Eq. (\ref{eqn:ActionEin}), the gravitational field 
is described by Einstein gravity, 
while transformed from the original action Eq. (\ref{eqn:ActionOrg}), 
and the extra scalar field $\phi$ is found from Eq. (\ref{eqn:OmegaPhiAsym}) 
to fall off faster than $\bar{\Omega}^{( n - 1 ) / 2}$ when the effective cosmological 
constant is subtracted. Therefore, Eq. (\ref{eqn:ConfNoetEqual}) holds 
even in the Einstein frame Eq. (\ref{eqn:ActionEin}), 
and we integrate it as $\BQcc[\xi] = \BQspI[\xi]$ by setting    
the integration constant so that it vanishes for the pure AdS spacetime.  
The mass $\bar{M}$ and the angular momenta $\BJI$ in the Einstein frame 
defined by Eq. (\ref{eqn:MassAngMomDefE}) are therefore found to be given by  
the Noether charges as  
\begin{equation} 
\bar{M} \equiv \BQspI[\BXiT] , ~~~~~ 
\BJI \equiv - \BQspI[\BXiI] .   
\label{eqn:MassAngMomNoetherE}  
\end{equation} 
Moreover, the entropy $\bar{S}$ in the Einstein frame is defined by the Noether charge 
as Eq. (\ref{eqn:EntropyEDef}), 
and hence we see that the first law of black hole thermodynamics 
is satisfied in the Einstein frame by the mass $\bar{M}$, the angular momenta $\BJI$, 
and the entropy $\bar{S}$ as 
\begin{equation} 
\frac{\bar{\kappa}}{2 \pi} \delta \bar{S} = \delta \bar{M} - \BOmI \delta \BJI .  
\label{eqn:FirstLawE} 
\end{equation} 

We further note that the value $\omega_0$ of the conformal factor $\omega$ 
at infinity, 
which depends only on the curvature length $\ell$ at infinity, 
does not change even when the mass and the angular momenta are varied. 
We thus find from Eqs. (\ref{eqn:MassAngMomRel}), (\ref{eqn:AngVelRel}), 
(\ref{eqn:SurfGravRel}), (\ref{eqn:EntropyEqual}), 
and (\ref{eqn:FirstLawE}), that the first law is satisfied also in the original frame as 
\begin{equation} 
\frac{\kappa}{2 \pi} \delta S = \delta M - \OmI \delta \JI . 
\label{eqn:FirstLawO} 
\end{equation} 

\section{Summary and discussion} 
\label{sec:ConcDis} 

We considered the class of gravitational theories described by the action 
Eq. (\ref{eqn:ActionOrg}), and proved first that conformal charges, i.e., 
well-defined conserved quantities constructed through conformal completion, 
in an asymptotically AdS spacetime remain untouched 
under the conformal transformation into the Einstein frame. 
We then considered a stationary and axisymmetric asymptotically AdS black hole, 
and defined the mass and the angular momenta of the black hole 
by these conformal charges as Eqs. (\ref{eqn:MassAngMomDefE}) 
and (\ref{eqn:MassAngMomDefO}) in both the original frame and the Einstein frame. 
Then, by taking into account the rescaling of the timelike Killing vector, 
the behavior of the thermodynamical variables under the conformal transformation 
into the Einstein frame was derived. In particular, we found that the entropies 
in the two frames coincide with each other, as in the asymptotically 
flat case \cite{JacobsonKM95}. 
By noticing that the result of Ref. \cite{HollandsIM} implies that 
the first law of black hole thermodynamics is satisfied in the Einstein frame, 
it was shown that the first law is satisfied also in the original frame, 
when the mass and the angular momenta are defined by the conformal charges 
and the entropy is defined by the Noether charge in the same manner as in the case 
of an asymptotically flat black hole. 
We also pointed out that it is naturally understood in the symplectic formalism 
that the mass satisfying the first law should be necessarily defined associated with 
the timelike Killing vector that is fixed with respect to ``the AdS background'', i.e., 
the universal structure determined by the boundary conditions at infinity. 

The action Eq. (\ref{eqn:ActionOrg}) describes 
Einstein gravity when $f(R) = R$. Hence, the result in this paper shows, 
in particular, that the first law of an asymptotically AdS black hole 
in Einstein gravity is generally satisfied 
if the mass and the angular momenta are defined by the conformal charges 
associated with the Killing vectors that are fixed with respect to ``the AdS background''. 
(See also Ref. \cite{PapaSke} for another derivation of the first law in Einstein gravity.) 
This is true also for higher curvature gravity within the class of 
Eq. (\ref{eqn:ActionOrg}). 
Actually, it is not difficult to generalize this result further. 
It is known \cite{LegendreTr} 
that the gravitational theories whose action is given by 
\begin{equation} 
I = \int d^n x \sqrt{- g} \left[ \frac{f( g^{a b} , R_{a b} )}{16 \pi G} + L_m \right] , 
\label{eqn:ActionL} 
\end{equation} 
where $f( g^{a b} , R_{a b} )$ is a nonsingular function of the metric $g^{a b}$ and 
the Ricci tensor $R_{a b}$, 
are transformed into the Einstein frame by the Legendre transformation 
\begin{equation} 
g^{a b} \rightarrow 
\bar{g}^{a b} \equiv \left[ - \det \frac{\partial ( 
\sqrt{- g} f(g^{c d} , R_{c d}))}{\partial R_{e f}} 
\right]^{- 1 / ( n - 2 )} \frac{\partial ( \sqrt{- g} f(g^{c d} , R_{c d}))}{\partial R_{a b}} .  
\label{eqn:GBarDef} 
\end{equation} 
In this case, the metric $\bar{g}^{a b}$ in the Einstein frame 
is not necessarily proportional to the metric $g^{a b}$ in the original frame, 
in contrast to the case of Eq. (\ref{eqn:ActionOrg}). 
However, $\bar{g}^{a b}$ is constructed only from $g^{a b}$ and $R_{a b}$. 
Moreover, $R_{a b}$ is proportional to $g_{a b}$ 
when the spacetime $( {\cal M} , g_{a b} )$ in the original frame 
is the pure AdS spacetime. Therefore, in the pure AdS spacetime, 
$\bar{g}^{a b}$ is proportional to $g^{a b}$, and hence 
the Legendre transformation Eq. (\ref{eqn:GBarDef}) 
reduces to a conformal transformation. 
This implies that in an asymptotically AdS spacetime, where the Ricci tensor $R_{a b}$ 
is asymptotically proportional to the metric $g_{a b}$, the Legendre transformation 
asymptotically approaches the conformal transformation. 
Thus, as we can expect from the result in Sec. \ref{sec:TransAndConsQ}, 
it will be possible to show that 
conformal charges in an asymptotically AdS spacetime 
are invariant under the transformation into the Einstein frame 
even in the case of Eq. (\ref{eqn:ActionL}). 
In addition, it has been proven \cite{KogaMaeda98} 
that the Noether charge $\QspH[\XiH]$ 
conjugate to the Killing vector $\XiH^a$, which gives the entropy of a black hole, 
also remains unchanged under the Legendre transformation. 
Then, we can employ the same method as shown in Sec. \ref{sec:FirstLaw} 
in order to prove that  
the mass $M$ and the angular momenta $\JI$ 
defined by Eq. (\ref{eqn:MassAngMomDefO}), 
along with the entropy $S$ defined by Eq. (\ref{eqn:EntropyODef}),  
satisfy the first law of an asymptotically AdS black hole 
even in the class of gravitational theories Eq. (\ref{eqn:ActionL}).   

However, in order to prove the first law 
for general theories of gravity,  
which may not be transformed into the Einstein frame, 
it actually looks more systematic   
to consider the Noether charges $\QspI[\XiT]$ and $\QspI[\XiI]$, 
rather than the conformal charges $\Qcc[\XiT]$ and $\Qcc[\XiI]$, 
and analyze the integrability condition of these Noether charges. 
In addition, the asymptotic condition on matter fields at infinity 
also should be analyzed in a systematic manner, 
while we simply assumed in this paper that they fall off rapidly enough. 

\acknowledgments
I like to thank N. Okuyama for discussions and K. Maeda for encouragement. 

\appendix 
\section{Four-dimensional Kerr--AdS black hole} 

The Kerr--AdS black hole is a solution in the class of gravitational theories 
Eq. (\ref{eqn:ActionOrg}) if there exists a negative constant $R_0$ that satisfies 
Eq. (\ref{eqn:RicciEq}), since the Ricci tensor $R_{a b}$ of the Kerr--AdS 
black hole spacetime takes the form of Eq. (\ref{eqn:RicciConstCurv}). 
Then, in this appendix, we calculate the mass and the angular momentum of 
the four-dimensional Kerr--AdS black hole 
in both of the original frame and the Einstein frame 
by using Eqs. (\ref{eqn:QExprOrg}) and (\ref{eqn:QExprEin}). 

We begin with the metric of the Kerr--AdS black hole 
in the coordinate system $( t', r, \theta, \phi' )$ that is {\it rotating} at infinity,  
\begin{equation} 
ds^2 = - \frac{\Delta}{\Sigma} \left[ d t' - \frac{a}{\Xi} \sin^2 \theta d \phi' \right]^2 
+ \frac{\Delta_{\theta} \sin^2 \theta}{\Sigma} 
\left[ \frac{r^2 + a^2}{\Xi} d \phi' - a d t' \right]^2 
+ \frac{\Sigma}{\Delta} d r^2 + \frac{\Sigma}{\Delta_{\theta}} d \theta^2 , 
\label{eqn:MetricRot} 
\end{equation} 
where 
\begin{equation} 
\Delta \equiv \left( r^2 + a^2 \right) \left( 1 + \frac{r^2}{l^2} \right) - 2 G m r , ~~~~ 
\Delta_{\theta} \equiv 1 - \frac{a^2}{l^2} \cos^2 \theta , ~~~~ 
\Sigma \equiv r^2 + a^2 \cos^2 \theta , ~~~~
\Xi \equiv 1 - \frac{a^2}{l^2} .  
\label{eqn:MFuncDef} 
\end{equation} 
We then transform Eq. (\ref{eqn:MetricRot}) into the nonrotating coordinate system 
$( t , r , \theta , \phi )$ by the coordinate transformation 
\begin{equation} 
t \equiv t' , ~~~~~ \phi \equiv \phi' + \frac{a}{l^2} t' , 
\label{eqn:CoorTransNRot} 
\end{equation} 
where the metric is now written as 
\begin{equation} 
ds^2 = - \frac{\Delta}{\Sigma \Xi^2} \left[ \Delta_{\theta} d t 
- a \sin^2\theta d \phi \right]^2 + \frac{\Delta_{\theta} \sin^2 \theta}{%
\Sigma \Xi^2} \left[ ( r^2 + a^2 ) d \phi - a \left( 1 + \frac{r^2}{l^2} \right) d t \right]^2 
+ \frac{\Sigma}{\Delta} d r^2 + \frac{\Sigma}{\Delta_{\theta}} d \theta^2 ,   
\label{eqn:MetricNRot} 
\end{equation} 
and the angular coordinates range as $0 < \theta < \pi$ and 
$0 \leq \phi < 2 \pi$. 
Here we note that the timelike Killing vector $\XiT^a$ 
and the axial Killing vector $\xi_{(\phi)}^a$ defined 
in the nonrotating coordinate system by 
\begin{equation} 
\XiT^a \equiv \left( \frac{\partial}{\partial t} \right)^a , ~~~~~ 
\xi_{(\phi)}^a \equiv \left( \frac{\partial}{\partial \phi} \right)^a
\end{equation} 
are expressed, respectively, in the rotating coordinate system as 
\begin{equation} 
\XiT^a 
= \left( \frac{\partial}{\partial t'} \right)^a 
- \frac{a}{l^2} \left( \frac{\partial}{\partial \phi'} \right)^a , ~~~~~ 
\xi_{(\phi)}^a 
= \left( \frac{\partial}{\partial \phi'} \right)^a  . 
\label{eqn:CoorTrTKilVec} 
\end{equation} 

Now we assume that the physical metric $g_{a b}$ in the original frame is 
given by Eq. (\ref{eqn:MetricNRot}). 
We then calculate the mass and the angular momentum of the black hole 
in the original frame from Eq. (\ref{eqn:QExprOrg}) with $n = 4$. 
Although a coordinate transformation of $r$ and $\theta$ is necessary 
in order to obtain the metric that behaves asymptotically as 
Eq. (\ref{eqn:StaticPureAdSO}), we can choose $r^{- 1}$ as the conformal factor 
$\Omega$ since $r^{- 1}$ is shown to vanish on ${\cal I}$. 
Thus, by transforming the radial coordinate from $r$ to $\Omega \equiv r^{- 1}$, 
we find that the unphysical metric $\hat{g}_{a b}$ in the original frame is given  
on ${\cal I}$ ($\Omega = 0$) as 
\begin{equation} 
d \hat{s}^2 = - \frac{\Delta_{\theta}}{l^2 \Xi} d t^2 
+ l^2 d \Omega^2 
+ \frac{1}{\Delta_{\theta}} d \theta^2 
+ \frac{1}{\Xi} \sin^2 \theta d \phi^2 ,   
\end{equation} 
and then the normal $\hat{n}^a$ to ${\cal I}$, the future-pointing timelike unit normal 
$\hat{N}^a$ to ${\cal C}$, and the volume element $\sqrt{\hat{\sigma}}$ of ${\cal C}$ 
are found to be given by 
\begin{equation} 
\hat{n}^a = \frac{1}{l^2} \left( \frac{\partial}{\partial \Omega} \right)^a , ~~~~ 
\hat{N}^a = l \sqrt{\frac{\Xi}{\Delta_{\theta}}} 
\left( \frac{\partial}{\partial t} \right)^a , ~~~~~ 
\sqrt{\hat{\sigma}} = \frac{1}{\sqrt{\Xi \Delta_{\theta}}} \sin \theta . 
\label{eqn:IntEntities} 
\end{equation} 
Since the relevant components of the Weyl tensor $C_{a b c d}$ associated with 
the physical metric $g_{a b}$ are calculated as 
\begin{eqnarray} 
C_{t r t r} & = & 
- \frac{G m r \Delta_{\theta}}{\Xi^2 \Sigma^3} \left[ 
2 \Delta_{\theta} 
+ \frac{a^2 \sin^2 \theta}{\Delta} \left( 1 + \frac{r^2}{l^2} \right)^2 \right] 
\left( r^2 - 3 a^2 \cos^2 \theta \right) , 
\label{eqn:WeylCompMass} \\ 
C_{\phi r t r} & = & 
\frac{G m a r \Delta_{\theta} \sin^2 \theta }{\Xi^2 \Sigma^3} \left[ 
2 + \frac{1}{\Delta} \left( 1 + \frac{r^2}{l^2} \right) 
\left( r^2 + a^2 \right) \right] 
\left( r^2 - 3 a^2 \cos^2 \theta \right) , 
\label{eqn:WeylCompAngMom} 
\end{eqnarray} 
we obtain the $t-t$ and $\phi-t$ components of the tensor $\hat{\cal E}_{a b}$ as 
\begin{equation} 
\hat{\cal E}_{t t} = - \frac{G m \Delta_{\theta}}{l^2 \Xi^2} \, 
\left[ 2 + \frac{a^2 }{l^2} \left( 1 - 3 \cos^2 \theta \right) \right] , ~~~~~ 
\hat{\cal E}_{\phi t} = \frac{G m a \Delta_{\theta}}{l^2 \Xi^2} \, 
\left[ 2 + \left( 1 - 3 \cos^2 \theta \right) \right] . 
\label{eqn:WeylElctrKAdS} 
\end{equation} 
From Eqs. (\ref{eqn:QExprOrg}), (\ref{eqn:ConfFactToEin}), (\ref{eqn:IntEntities}), 
and (\ref{eqn:WeylElctrKAdS}), we see that the mass $M$ 
and the angular momentum $J$ in the original frame, 
which are defined by Eq. (\ref{eqn:MassAngMomDefO}),  
are given as  
\begin{equation} 
M \equiv \Qcc[\XiT] = \omega_0^2 \frac{m}{\Xi^2} , ~~~~~ 
J \equiv - \Qcc[\xi_{( \phi )}] = \omega_0^2 \frac{m a}{\Xi^2} .  
\label{eqn:MassAngMomKAdSO} 
\end{equation} 

To calculate the mass and the angular momentum in the Einstein frame, 
we notice here that the conformal factor $\omega$ is 
constant in this case, which we thus denote as $\omega_0$, since 
the Ricci scalar $R$ of the Kerr--AdS black hole spacetime is constant.  
The physical metric $\bar{g}_{a b}$ in the Einstein frame is then found to be given 
by the same form as Eq. (\ref{eqn:MetricNRot}), 
while $t$, $r$, $a$, $\ell$, and $m$ are replaced by  
\begin{equation} 
\bar{t} \equiv \omega_0 t , ~~~ \bar{r} \equiv \omega_0 r, ~~~ 
\bar{a} \equiv \omega_0 a , ~~~ \bar{\ell} \equiv \omega_0 \ell , ~~~ 
\bar{m} \equiv \omega_0 m , 
\label{eqn:KAdSRescale} 
\end{equation} 
respectively, 
which is understood from the rescaling property Eq. (\ref{eqn:Rescaling}) at infinity. 
Then, $\tilde{n}^a$, $\tilde{N}^a$, $\sqrt{\tilde{\sigma}}$, and $\tilde{\cal E}_{a b}$, 
which appear in Eq. (\ref{eqn:QExprEin}),  
take the same forms as those given in Eqs. (\ref{eqn:IntEntities}) 
and (\ref{eqn:WeylElctrKAdS}), but with the unbarred quantities replaced by the barred 
ones. The mass $\bar{M}$ and the angular momentum $\bar{J}$ in the Einstein 
frame defined by Eq. (\ref{eqn:MassAngMomDefE}) are therefore 
computed from Eqs. (\ref{eqn:QExprEin}) and (\ref{eqn:KAdSRescale}) as 
\begin{equation} 
\bar{M} \equiv \BQcc[\BXiT] = \frac{\bar{m}}{\Xi^2} = \omega_0 \frac{m}{\Xi^2}, ~~~~~ 
\bar{J} \equiv - \BQcc[\bar{\xi}_{( \phi )}] = \frac{\bar{m} \bar{a}}{\Xi^2} 
= \omega_0^2 \frac{m a}{\Xi^2} .  
\label{eqn:MassAngMomKAdSE} 
\end{equation} 
We thus see that the mass and the angular momentum 
of the four-dimensional Kerr--AdS black hole in Einstein gravity 
\cite{GibbonsPP-,CaldarelliCK00} are correctly reproduced, i.e., $\omega_0 = 1$ 
in Eqs. (\ref{eqn:MassAngMomKAdSO}) and (\ref{eqn:MassAngMomKAdSE}) 
or the barred expression in Eq. (\ref{eqn:MassAngMomKAdSE}), and also that 
Eq. (\ref{eqn:MassAngMomRel}) is indeed satisfied. 

Finally, we note that the sign of the second equation in 
Eq. (\ref{eqn:MassAngMomDefE}) and 
the sign of the second term of the first equation in Eq. (\ref{eqn:CoorTrTKilVec}) 
are opposite to those presented in Ref. \cite{GibbonsPP-}. 
However, these two signs cancel with each other in the calculation 
performed in Ref. \cite{GibbonsPP-}, and hence the eventual conclusion 
that the mass defined by Eq. (\ref{eqn:MassAngMomDefE}) satisfies 
the first law of black hole thermodynamics is unaltered. 



\begin{thebibliography}{99} 

\bibitem{BardeenCH73} 
J. M. Bardeen, B. Carter, and S. W. Hawking, Commun. Math. Phys. \textbf{31}, 161 (1973). 
\bibitem{GibbonsPP-} 
G. W. Gibbons, M. J. Perry, and C. N. Pope, hep-th/0408217. 
\bibitem{AdSCFT} 
J. Maldacena, 
Adv. Theor. Math. Phys. {\bf 2}, 231 (1998); 
S. S. Gubser, I.R. Klebanov and A. M. Polyakov, 
Phys. Lett. {\bf B428}, 105 (1998); 
E. Witten, Adv. Theor. Math. Phys. {\bf 2}, 253 (1998); 
O. Aharony, S. S. Gubser, J. Maldacena, H. Ooguri and Y. Oz, 
Phys. Rep. {\bf 323}, 183 (1998). 
\bibitem{CallanFMP85} 
C. G. Callan, D. Friedan, E. J. Martinec, and M. J. Perry, Nucl. Phys. \textbf{B262}, 593 (1985). 
\bibitem{BirrellDavies82} 
N. D. Birrell and P. C. W. Davies, 
\textit{Quantum Fields in Curved Space} (Cambridge University Press, Cambridge, England, 1982). 
\bibitem{FirstLawAF} 
R. M. Wald, Phys. Rev. D \textbf{48}, 3427 (1993); 
V. Iyer and R. M. Wald, Phys. Rev. D \textbf{50}, 846 (1994); 
T. Jacobson, G. Kang, and R. C. Myers, Phys. Rev. D \textbf{49}, 6587 (1994).  
\bibitem{RaczWald} 
I. Racz and R. M. Wald, Class. Quantum Grav. \textbf{9}, 2643 (1992); 
Class. Quantum Grav. \textbf{13}, 539 (1996). 
\bibitem{AshtekarMagnonDas}
A. Ashtekar and A. Magnon,
Class.\ Quantum Grav. \textbf{1}, 39 (1984); 
A.\ Ashtekar and S.\ Das,
Class.\ Quantum Grav. \textbf{17}, 17 (2000). 
\bibitem{HollandsIM} 
S. Hollands, A. Ishibashi, and D. Marolf, hep-th/0503045. 
\bibitem{WaldZoupas00} 
R. M. Wald and A. Zoupas, Phys. Rev. D \textbf{61}, 084027 (2000). 
\bibitem{OkuyamaKoga05} 
N. Okuyama and J. Koga, Phys. Rev. D \textbf{71}, 084009 (2005). 
\bibitem{NojiriOdintsov02} 
S. Nojiri and S. D. Odintsov, Phys. Rev. D \textbf{66}, 044012 (2002). 
\bibitem{JacobsonKM95} 
T. Jacobson, G. Kang, and R. C. Myers, Phys. Rev. D \textbf{52}, 3518 (1995). 
\bibitem{KogaMaeda98} 
J. Koga and K. Maeda, Phys. Rev. D \textbf{58}, 064020 (1998). 
\bibitem{PapaSke} 
I. Papadimitriou and K. Skenderis, hep-th/0505190. 
\bibitem{LegendreTr} 
G. Magnano, M. Ferraris, and M. Francaviglia, Gen. Relariv. Gravit. 
\textbf{19}, 465 (1987); 
M. Ferraris, M. Francaviglia, and G. Magnano, Class. Quantum Grav. \textbf{7}, 261 (1990); 
A. Jakubiec and J. Kijowski, Phys. Rev. D \textbf{37}, 1406 (1998). 
\bibitem{CaldarelliCK00} 
M. M. Caldarelli, G. Cognola, and D. Klemm, Class. Quantum Grav. \textbf{17}, 399 (2000). 
\end{thebibliography}
\end{document}